\documentclass[useAMS,usenatbib,usemath,usegraphicx,usedcolumn]{mn2e}  

\newcommand{\kms}{km\ s$^{-1}$~}

\newcommand{\hinv}{h$^{-1}$}
\newcommand{\mhinv}{\rm{h^{-1}}}

\def\aj{{AJ}}                   
\def\apj{{ApJ}}                 
\def\apjl{{ApJ}}                
\def\aap{{A\&A}}                
\def\mnras{{MNRAS}}             

\title[The Environments of SLACS Gravitational Lenses]{The Environments of SLACS Gravitational Lenses}

\author[M. W. Auger]{M. W. Auger$^{1}$\thanks{E-mail: mauger@physics.ucdavis.edu}\\
$^{1}$
   Department of Physics, University of California, 1 Shields Avenue,
   Davis, CA 95616, USA }

\begin{document}
\date{}
\pagerange{\pageref{firstpage}--\pageref{lastpage}} \pubyear{2002}

\maketitle

\label{firstpage}

\begin{abstract}
We report on an investigation of the environments of the SLACS sample of gravitational lenses. The local and global environments of the lenses are characterized using SDSS photometry and, when available, spectroscopy. We find that the lens systems that are best modelled with steeper than isothermal density profiles are more likely to have close companions than lenses with shallower than isothermal profiles. This suggests that the profile steepening may be caused by interactions with a companion galaxy as indicated by N-body simulations of group galaxies. The global environments of the SLACS lenses are typical of non-lensing SDSS galaxies with comparable properties to the lenses, and the richnesses of the lens groups are not as strongly correlated with the lens density profiles as the local environments. Furthermore, we investigate the possibility of line-of-sight contamination affecting the lens models but do not find a significant over-density of sources compared to lines of sight without lenses.
\end{abstract}


\section{INTRODUCTION}
The Sloan Lens ACS Survey \citep[SLACS;][]{bolton} is a large sample of strong gravitational lenses derived from the Sloan Digital Sky Survey (SDSS). The lens sample has proved to be particularly useful due to the high quality of the data; all of the SLACS lenses have known lens and source redshifts, stellar velocity dispersions for the lensing galaxies have been measured from the SDSS spectra, all of the systems have {\em Hubble Space Telescope} ({\em HST}) ACS imaging in two bands for accurate non-parametric lens modelling, and all of the sources are extended and therefore provide additional constraints on the mass profile of the lensing galaxy. Furthermore, the density of background sources in the {\em HST} imaging allows a weak lensing analysis of the ensemble sample of lenses \citep{gavazzi}; the major drawback of the lens sample is that it is unlikely to find any variable sources that would provide time delays.

\citet{koopmansSLACS} have used the SLACS lenses to show that early-type galaxies have isothermal total inner density profiles with very little intrinsic scatter (approximately 6 per cent) assuming a uniformity in the environments of the lenses that has not been rigorously tested.  While the SLACS lenses seem to lie on the Fundamental Plane \citep{bolton,treu,bolton07} and do not differ noticeably from other SDSS galaxies with similar luminosities and stellar velocity dispersions, the local environments of the lenses might affect the mass profiles of the lensing galaxies \citep[e.g.,][]{rusin,dobke,augerb}. If some of the lenses are being perturbed by neighbouring galaxies the intrinsic scatter of the density slope for {\em isolated} early-type galaxies might be even smaller than 6 per cent. Furthermore, it has been suggested that line-of-sight (LOS) contamination significantly affects the SLACS lenses \citep{guimaraes} and the density slope might be expected to be shallower than originally reported.

We report on a spectroscopic and photometric evaluation of the environments and lines of sight of the 15 SLACS lenses investigated by \citet{koopmansSLACS}. A weighting scheme is used to determine the effective number of potential perturbing companions to each lens galaxy. We also characterize the `richness' of the global environment of each lens field and quantify the number of galaxies along the LOS to the lens. Throughout this paper the term `global environment' is used to describe the group, cluster, or field in which the lens resides while the `local environment' describes the environment within $\approx 100$ \hinv kpc of the lensing galaxy. A $\Lambda$CDM cosmology with $\Omega_M = 0.27$ and $\Omega_\Lambda = 0.73$ is used to determine all physical distances, which are measured in \hinv units.

\section{DATA SAMPLES}
\subsection{Lens Systems}
Our primary intent is to investigate a correlation between the inner density profile of the lensing galaxy and the galaxy's local environment. We therefore limit our sample to the subset of SLACS lenses with published inner density slopes. This sample consists of 15 early-type galaxies with lens redshifts between $z = 0.063$ and $z = 0.332$ and stellar velocity dispersions from $\sigma = 178$ \kms to $\sigma = 330$ \kms \citep{koopmansSLACS}. The density slopes range from $\alpha = 1.82$ to $\alpha = 2.34$ where $\rho \propto r^{-\alpha}$, and the typical error on $\alpha$ for a given system is approximately 5 per cent \citep{koopmansSLACS}. Four of the systems have shallower than isothermal density slopes (i.e., $\alpha < 2$), six are approximately consistent with isothermal, and five lenses have steeper than isothermal profiles. The lens characteristics are summarized in Table \ref{table_lens_properties}.

\begin{table}
 \begin{center}
  \caption{Properties of the 15 SLACS gravitational lenses investigated in this work. Columns: (1) lens name; (2) lens redshift; (3) lens stellar velocity dispersion; (4) lens mass slope; (5) photometric estimate for the number of close companions; (6) proxy for the richness of the local environment.}
  \begin{tabular}{@{}ld{2}cd{2}d{2}d{1}@{}}
  \hline
  \multicolumn{1}{l}{Name$^\dagger$} & \multicolumn{1}{r}{$z^\dagger$} & \multicolumn{1}{r}{$\sigma_{\rm SDSS}^\dagger$} & \multicolumn{1}{r}{$\alpha^\dagger$} & \multicolumn{1}{r}{$N_w$} & \multicolumn{1}{r}{$R$} \\ 
 \hline
 SDSSJ0037-0942  &   0.1954  &  265  & 2.05  & 0.11  &   3.2 \\
 SDSSJ0216-0813  &   0.3317  &  332  & 2.05  & 0.02  &   1.4 \\
 SDSSJ0737+3216  &   0.3223  &  310  & 2.34  & 0.86  &   2.3 \\
 SDSSJ0912+0029  &   0.1642  &  313  & 1.82  & 0.03  &   3.7 \\
 SDSSJ0956+5100  &   0.2405  &  299  & 2.04  & 1.98  &   8.2 \\
 SDSSJ0959+0410  &   0.1260  &  212  & 2.18  & 0.80  &  22.5 \\
 SDSSJ1250+0523  &   0.2318  &  254  & 2.26  & 0.01  &   6.1 \\
 SDSSJ1330-0148  &   0.0808  &  178  & 2.18  & 5.80  &  87.1 \\
 SDSSJ1402+6321  &   0.2046  &  275  & 1.95  & 0.24  &   2.8 \\
 SDSSJ1420+6019  &   0.0629  &  194  & 2.03  & 0.00  &   4.8 \\
 SDSSJ1627-0053  &   0.2076  &  275  & 2.21  & 0.96  &   7.1 \\
 SDSSJ1630+4520  &   0.2479  &  260  & 1.85  & 0.00  &   4.1 \\
 SDSSJ2300+0022  &   0.2285  &  283  & 2.07  & 0.91  &   7.2 \\
 SDSSJ2303+1422  &   0.1553  &  260  & 1.82  & 0.00  &   2.9 \\
 SDSSJ2321-0939  &   0.0819  &  236  & 1.87  & 0.05  &   4.0 \\
\hline
\label{table_lens_properties}
\end{tabular}
\end{center}
$\dagger$ Data from \citet{koopmansSLACS}.
\end{table}

\subsection{Lens Environments}
Photometric and spectroscopic data from the SDSS Data Release 6 \citep{sdss} are used to quantify the environments of the lensing galaxies. Catalogs are made of all SDSS primary galaxies with $r < 22.1$ mag that lie within 1.2 \hinv Mpc of the lensing galaxy. These catalogues include the five band SDSS photometry, photometric redshifts from the {\tt photoz} table of the SDSS catalogue, and spectroscopic redshifts for those galaxies targeted by SDSS. Due to the range in redshifts spanned by the lenses, the spectroscopic completeness varies significantly from system to system. Only the systems with $z \la 0.1$ contain sufficient spectroscopic redshifts to adequately characterize the global environments of the lenses.

\subsubsection{Local Environment}
\begin{figure}
 \begin{center}
 \includegraphics[width=0.45\textwidth]{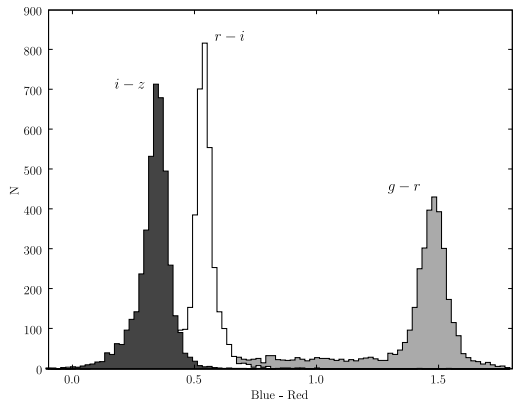}
 \caption{Distributions of colours for galaxies at $z = 0.25$ in the SDSS spectroscopic database. The dark gray distribution is $i-z$, the white distribution is $r-i$, and the light gray distribution is $g-r$. Note that there is a blue tail associated with each distribution, and the $g-r$ distribution is particularly affected. These tails are due to late-type galaxies and have been clipped when determining the means and widths of the distributions.}
 \label{figure_color_distribution}
 \end{center}
\end{figure}

We use the colours of the field galaxies and their offsets from the lens to determine the likelihood that a neighbour galaxy is physically associated with the lens. Only galaxies that are less than one magnitude brighter or 2.5 magnitudes fainter than the lens are used in the analysis. Fiducial colours of galaxies at the lens redshift are determined by querying all spectroscopic galaxies in the SDSS database that have redshifts within 1200 \kms of the redshift of the lensing galaxy (this is equivalent to $\Delta z \approx 0.004$ at the redshifts of the lenses). These empirical colour distributions are used to determine Gaussian weights for each of the field galaxies, which are also weighted by their distance from the lens system. The colour distributions for a $z = 0.25$ galaxy are shown in Figure \ref{figure_color_distribution}; note that there are long tails at the blue end of each distribution. These tails are due to late-type galaxies and are clipped from our analysis when determining the location and width of the distributions. The clipping biases us against finding late-type companion galaxies; however, this is not a strong bias because early-type galaxies are known to cluster much more strongly than late-type galaxies \citep{menaux}.

The final weighting scheme used is
\[
w_{gal} = f(d) \prod_{c} e^{-\frac{\Delta_{c}^{2}}{2 \sigma_{c}^{2}}},
\]
where $w_{gal}$ is the weight for each galaxy and $f(d)$ is given by
\[
f(d) =
\left\{
\begin{array}{cl}
 1.  &  d < 90~\mhinv kpc \\
 e^{-\frac{d - 90}{d_0}}  &  d > 90~\mhinv kpc
\end{array}
\right.
\]
where $d$ is the distance of the galaxy from the lens and $d_0$ is set to 20 \hinv kpc for an effective distance of approximately 100 \hinv kpc \citep[i.e. several truncation radii,][]{limousin}. The product is taken over the set of colours $c =\{g-r, r-i, i-z\}$, $\Delta_{c}$ is the difference between the mean of the empirical colour distribution and the measured colour of the galaxy, and $\sigma_{c}$ is the quadrature sum of the width of the colour distribution and the errors on the SDSS photometry. The $u$ filter is not used in our analysis due to its poor sensitivity and all of the galaxies in our sample have the three color photometry $\{g-r, r-i, i-z\}$ available from the SDSS database. We sum the weights of all of the galaxies in the field to determine an effective number of galaxies physically associated with the lensing galaxy, $N_w$. The photometric weighting is not equivalent to determining a photometric redshift for each galaxy; photometric redshifts characterize the most likely redshift of a source while our weights are a proxy for the likelihood of a source being at a given redshift (though the full probability distribution from a photometric redshift analysis could approximately provide the same information).

Our weighting algorithm was tested on low-redshift ($0.025 < z < 0.075$) SDSS galaxies that have luminosities and stellar velocity dispersions that are comparable to the SLACS lenses. These low-redshift galaxies have nearly complete spectroscopy for all neighbouring field galaxies with $r_{field} < r_{gal} + 2.5$. The weighting algorithm was employed as above, and a comparison was made between $N_w$ and the number of spectroscopically confirmed neighbours. $N_w$ correctly estimated the {\em total} number of companions within 100 \hinv kpc of the target galaxy for 64 per cent of the systems, where $N_w$ was rounded to the nearest integer to make the comparison. For the incorrectly matched systems, $N_w$ does not show a bias for over or underestimating the number of companions; 18 per cent were incorrectly matched in either case. However, $N_w$ is more robust as a binary indicator of whether there are any companions or not. Our simulations show that 67 per cent of systems with $N_w > 0.75$ have at least one spectroscopically confirmed companion, while 76 per cent of the systems with $N_w < 0.75$ do not have any companions. We suspect that many of the incorrect associations are due to late-type galaxies that are not identified well by our colour distributions.

\begin{figure}
 \begin{center}
 \includegraphics[width=0.45\textwidth]{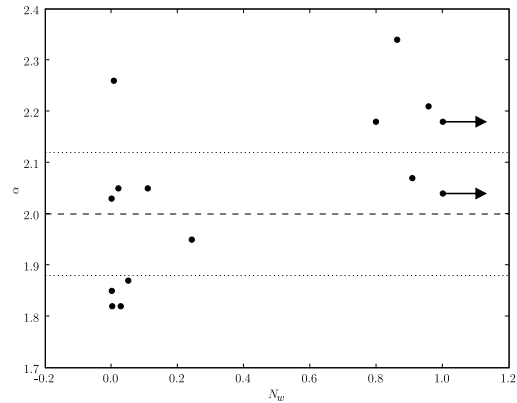}
 \caption{The number of photometrically estimated companions ($N_w$) versus the lens density slope ($\alpha$) inferred from lensing plus stellar dynamics models for 15 SLACS lenses. The two systems with $N_w > 1$ have been set to 1 to emphasize the bimodality. The dashed line at $\alpha = 2$ indicates an isothermal density profile and the dotted lines are 1-$\sigma$ deviations from isothermal.}
 \label{figure_lens_neighbors}
 \end{center}
\end{figure}

\begin{figure}
 \begin{center}
 \includegraphics[width=0.45\textwidth]{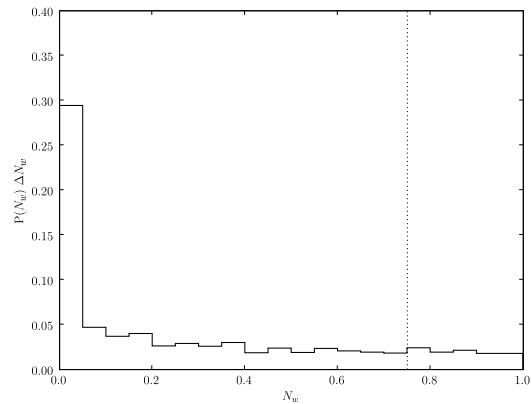}
  \caption{The distribution of $N_w$ for all of the SDSS comparison fields. Approximately 33 per cent of the fields have $N_w > 0.75$ (indicated by the vertical dotted line); this is slightly lower than the 40 per cent companion rate found for SLACS lenses.}
 \label{figure_comp_neighbors}
 \end{center}
\end{figure}

$N_w$ for each lens system is recorded in Table \ref{table_lens_properties}. There is a noticeable trend for lenses with steeper density profiles to have higher values of $N_w$ (Figure \ref{figure_lens_neighbors}). Furthermore, none of the shallower than isothermal lenses are found to be associated with a companion galaxy. A simple linear regression of the data (after setting $N_w = 1$ for the systems with $N_w > 1$) finds that the trend is real with 98.3 per cent confidence. A comparison sample of non-lens fields was made for each lens system to provide an external comparison. For each lens, the same analysis as described above was performed on 200 SDSS galaxies at the same redshift as the SLACS lens and with a measured stellar velocity dispersion within 20 \kms of the lens velocity dispersion. In some cases the number of comparison fields meeting these criteria is less than 200, in which case we used all fields that do meet the criteria. The composite distribution of $N_w$ for all of the comparison samples is shown in Figure \ref{figure_comp_neighbors}; a Kolmogorov-Smirnov (K-S) test was unable to distinguish between the distributions.

\subsubsection{Global Environment}
\begin{figure}
 \begin{center}
 \includegraphics[width=0.45\textwidth]{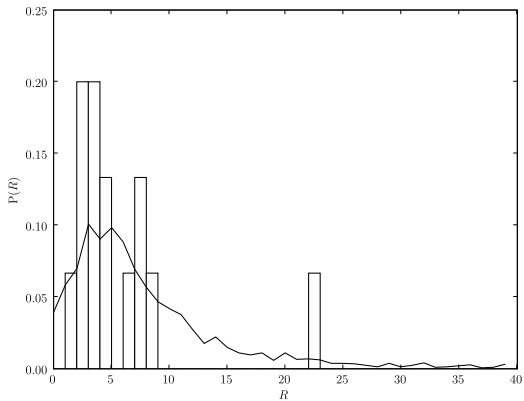}
  \caption{The distribution of $R$ for all of the SDSS comparison fields (black line) and 14 of the SLACS lenses. The lenses largely have richnesses near the peak of the distribution at $R \sim 5$. One lens, SDSSJ1330-0148, is in a cluster with $R = 87$ and has been omitted for clarity.}
 \label{figure_comp_rich}
 \end{center}
\end{figure}

A proper characterization of the global environment of a lens system would determine the velocity dispersion for the group that the lens resides in as well as the distance of the lens from the centre of the group. Neither of these can be quantified without spectroscopically defining the group membership; we are only able to perform this analysis for the lowest redshift lenses (Section \ref{section_spectroscopy}). Nevertheless, we attempt to characterize the `richness' of the global environment by employing a weighting scheme similar to the one used for the local environment. The same colour distributions are used and the analysis only differs by the weight term used to penalize the radial offset from the lens; the distance weighting for the global environment is
\[
f(d) = \frac{1}{\frac{d}{d_0} + 1}
\]
with $d_0$ set to 350 \hinv kpc, approximately the virial radius for moderate redshift groups \citep[e.g.,][]{augera}. The group richness, $R$, is the sum of all of the individual weights and is recorded in Table \ref{table_lens_properties}. While $N_w$ can be interpreted approximately as the number of neighbouring galaxies for each lens, $R$ should not be interpreted as the number of galaxies in the group, though it is a proxy for the density of early-type galaxies at a particular redshift. Figure \ref{figure_comp_rich} shows the distribution of richness for the SLACS lenses and the SDSS comparison sample; a K-S test cannot distinguish between the distributions. The lenses lie in typical environments and there is not a strong correlation between the global richness and the density slope for the SLACS lenses.

\subsubsection{Spectroscopic Sample}
\label{section_spectroscopy}
Three of the systems, SDSSJ1330-0148, SDSSJ1420+6019, and SDSSJ2321-0939, are at lower redshifts than the other lenses and their fields have nearly complete SDSS spectroscopy down to $r = r_{lens} + 2.5$. We use the SDSS redshifts to determine group properties for these lens fields. There is one spectroscopically identified companion within 100 \hinv kpc of SDSSJ1330-0148 but no spectroscopic neighbours are found for the other two systems, tentatively confirming the binary interpretation of $N_w$ for these three systems. However, due to the inability to closely pack fibres on the SDSS spectrograph, the local environments of these lenses are not completely probed spectroscopically. One of the systems, SDSSJ1420+6019, appears to be isolated and no group is found to be associated with the lens. For the remaining two lens systems, we determine the group velocity dispersion using the biweight estimate of the velocity distribution \citep{beers}, and the lens offset is determined with respect to the median position of all identified group members (Table \ref{table_groups}).

\begin{table}
 \begin{center}
  \caption{Group properties for the low-redshift SLACS lenses that lie in spectroscopically confirmed groups. Columns: (1) lens name; (2) number of spectroscopically confirmed group members; (3) distance from the lensing galaxy to the group center; (4) group velocity dispersion.}
  \begin{tabular}{@{}lrrr@{}}
  \hline
  Name &  \multicolumn{1}{c}{$N_{grp}$} & \multicolumn{1}{c}{Offset ($\arcsec$)} & \multicolumn{1}{c}{$\sigma$ (\kms)} \\
 \hline
SDSSJ1330-0148  &  87  &  167  &  905$\pm84$ \\
SDSSJ2321-0939  &  13  &   41  &  411$\pm58$ \\
\hline
\label{table_groups}
\end{tabular}
\end{center}
\end{table}

\subsubsection{Line-of-Sight Contamination}
The LOS `contamination' for a lens can be quantified by summing the number of objects within an aperture around the lensing galaxy. We use the SDSS photometric redshift catalogue to exclude local galaxies with redshifts $z < 0.001$ that do not strongly influence the lensing. We then count all background galaxies with offsets less than 30 arcsec from the lens and compare this sum with background galaxy counts in non-lens fields. The SDSS comparison sample for each lens contains 200 galaxies at the same redshift as the lens and with approximately the same luminosity and velocity dispersion. The mean, median, and standard deviation of the comparison samples are tabulated with the lens data in Table \ref{table_lineofsight}. The most significant deviation between the lenses and comparison fields is less than 1.5 $\sigma$ and there is not a systematic offset. Note that our comparison is between the LOS of a SLACS lens and the lines of sight to other SDSS massive early-type galaxies; random lines of sight might be more or less dense.

\begin{table}
 \begin{center}
  \caption{The number of non-local galaxies along the lines of sight to SLACS lenses compared to similar, non-lensing SDSS galaxies. Columns: (1) lens name; (2) number of galaxies along LOS to the lens; (3) mean number of galaxies along LOS to non-lensing SDSS galaxies; (4) median number of galaxies along LOS to non-lensing SDSS galaxies; (5) standard deviation of number of galaxies along LOS to non-lensing SDSS galaxies.}
  \begin{tabular}{@{}lrrrr@{}}
  \hline
  &  &  \multicolumn{3}{c}{Comparison Fields} \\
  Name & $N_{lens}$ & $N_{mean}$ & $N_{med}$ & $\sigma_N$ \\
 \hline
 SDSSJ0037-0942  &   5  &   5.4  &   5.0  &   3.2  \\
 SDSSJ0216-0813  &  11  &   8.0  &   7.0  &   4.6  \\
 SDSSJ0737+3216  &   6  &   7.3  &   7.0  &   4.0  \\
 SDSSJ0912+0029  &   9  &   7.3  &   7.0  &   3.6  \\
 SDSSJ0956+5100  &  10  &   7.7  &   7.0  &   3.9  \\
 SDSSJ0959+0410  &   5  &   4.1  &   4.0  &   2.4  \\
 SDSSJ1250+0523  &   5  &   5.9  &   5.0  &   3.2  \\
 SDSSJ1330-0148  &   5  &   3.8  &   3.0  &   2.5  \\
 SDSSJ1402+6321  &   6  &   6.0  &   5.0  &   3.2  \\
 SDSSJ1420+6019  &   3  &   5.1  &   4.0  &   3.6  \\
 SDSSJ1627-0053  &   9  &   6.0  &   6.0  &   3.1  \\
 SDSSJ1630+4520  &   4  &   6.0  &   6.0  &   3.3  \\
 SDSSJ2300+0022  &  11  &   6.4  &   6.0  &   3.4  \\
 SDSSJ2303+1422  &   6  &   5.2  &   5.0  &   3.0  \\
 SDSSJ2321-0939  &   9  &   5.2  &   5.0  &   3.2  \\
\hline
\label{table_lineofsight}
\end{tabular}
\end{center}
\end{table}

\section{DISCUSSION}
We find that the global environments of the SLACS lenses are typical of other massive early-type galaxies found in the SDSS. Two of the steeper than isothermal systems lie in very over-dense regions but the remaining lenses all have richness values that fall near the peak of the richness distribution (Figure \ref{figure_comp_rich}). Furthermore, the suggestion that the SLACS lenses are affected by LOS contamination does not seem to be merited by the data. While there are other galaxies along the lines of sight to the lens systems, the LOS densities do not significantly deviate from the densities along comparable lines of sight. We therefore expect that the parameter estimates should not be affected as proposed by \citet{guimaraes}.

The SDSS photometric data indicate that the SLACS lenses are only slightly more likely to be associated with companion galaxies than comparable lenses selected from the SDSS, though the uncertainty of the photometric identifications makes the difference negligible. We therefore conclude that SLACS lenses do lie in typical environments both globally and locally. However, we also find that lens systems with steeper than isothermal density slopes are preferentially associated with companion galaxies compared to lenses with shallower density slopes. N-body simulations suggest that interactions with neighbouring galaxies can induce a steepening in the density slope \citep{dobke} and there are other lens systems with companion galaxies that are found to be best modelled with steeper than isothermal profiles \citep[e.g.,][]{rusin,augerb}. The interaction-induced steepening is a transient effect and the density profile of the galaxy will return to isothermal approximately 0.5-2 Gyr after the encounter with the neighbour \citep{dobke}. This may account for the large range of $N_w$ for lenses with nearly isothermal profiles; isothermal lenses with a companion may be in the relaxed state before or after an encounter with the neighbour galaxy. This stripping mechanism may also account for local observations of dark matter deficient galaxies \citep[e.g.,][]{romanowsky,proctor}.

We note that one steeper-profile system, SDSSJ1250+0523, is not photometrically associated with any neighbouring galaxies; this perhaps illustrates the limits of using photometry to find perturbing companions or demonstrates that other factors also influence the slope of the density profile. Additionally, we have not found an environmental bias to account for the shallower lenses. However, if a lens galaxy is embedded in a cluster, the joint profile of the cluster and galaxy would tend to be modelled with a shallower than isothermal profile if a single-component power law is used (ignoring interactions between the two halos). This effect is dependent on the location of the lens with respect to the centre of the cluster, which our photometric analysis is unable to address. More complete field spectroscopy would better characterize the global environments of these lens systems and allow correlations between the mass slopes and cluster centre offsets to be investigated. Furthermore, a spectroscopic investigation of the local environments of the complete sample of SLACS early-type lenses would confirm the correlation indicated by our photometric analysis and provide strong evidence for truncation caused by galaxy interactions.


\section*{Acknowledgements}
The author would like to thank Chris Fassnacht and Ami Choi, as well as the anonymous referee, for providing useful comments on the manuscript. This work has made extensive use of the SDSS database. Funding for the SDSS and SDSS-II has been provided by the Alfred P. Sloan Foundation, the Participating Institutions, the National Science Foundation, the U.S. Department of Energy, the National Aeronautics and Space Administration, the Japanese Monbukagakusho, the Max Planck Society, and the Higher Education Funding Council for England. Part of this work was supported by the European Community's Sixth Framework Marie Curie Research Training Network Programme, Contract No. MRTN-CT-2004-505183 ``ANGLES".


\label{lastpage}

\end{document}